\documentclass{aa}  
\usepackage{multirow}
\usepackage{graphicx}
\usepackage{txfonts}
\usepackage{lipsum}
\usepackage{subcaption} 
\usepackage{lscape} 
\usepackage{placeins}
\usepackage{xcolor}
\usepackage{hyperref}
\usepackage{comment}
\usepackage{booktabs}
\usepackage{threeparttable}
\begin{document}

\title{Formation of intermediate-mass black holes in young massive clusters detected with JWST: analytic mass estimates}

\titlerunning{Formation of intermediate-mass black holes in JWST clusters}

   \author{Viola Bocchi\inst{1}
        \and Mat\'ias Liempi\inst{1}
        \and  Dominik R.G. Schleicher\inst{1}
        }

\institute{Dipartimento di Fisica, Sapienza Universit\`a di Roma, Piazzale Aldo Moro 5, 00185 Rome, Italy\\  \email{bocchi.2109451@studenti.uniroma1.it}}

   \date{Received September 30, 20XX}

  \abstract
   {The James Webb Space Telescope (JWST) has revealed a population of dense stellar systems at high redshift, including the "Cosmic Gems" arc ($z \sim 10.2$) and the "Firefly Sparkle" ($z \sim 8.3$). With masses in the range of $10^5$~M$_\odot$-$10^7$~M$_\odot$ and half-mass radii in the range from $\sim0.4$-$15$~pc, these systems are ideally suited to form intermediate-mass black holes (IMBHs) via collision-based models. While direct N-body simulations are unfeasible for such a large population and given the high masses in many of the clusters, we estimate the IMBH masses  formed via runaway stellar collisions in these specific environments utilizing a  Fokker-Planck model together with an analytical framework for runaway collisions and mass loss through winds, which has been validated against direct N-body  simulations of compact star clusters. We apply this model to a sample of massive high-redshift clusters observed with JWST. Our estimates yield typical IMBH masses in the range of  $\sim10^2$~M$_\odot$  up to $\sim4\times 10^3$~M$_\odot$, implying typical formation efficiencies on the few percent level. The extreme compactness of the Cosmic Gems clusters ($R_h \sim 1$ pc) facilitates the formation of black hole seeds with high masses of $1600-2700 \, {\rm M}_\odot$. Low metallicity ($Z \lesssim 0.02 \, {\rm Z}_\odot$) is a critical factor for retaining the seed mass against stellar winds. We further demonstrate that the efficiencies obtained here are consistent with expectations based on direct N-body simulations.  Our results suggest that these dense, metal-poor clusters are viable factories for heavy seeds, capable of growing into the supermassive black holes observed in the early Universe. }


   \maketitle
\nolinenumbers
\section{Introduction}

The James Webb Space Telescope (JWST)\footnote{JWST: https://science.nasa.gov/mission/webb/} has meanwhile provided abundant results pointing to very massive and dense systems at high redshift, including systems with high efficiencies of star formation. Some of the first galaxies that were detected initially were even considered to be more massive than they should be within the $\Lambda$CDM framework  \citep[e.g.,][]{Labbe2023}, fact that was later attributed and explained through the active galactic nuclei (AGN) contamination of these sources \citep{Chworosky2024}. While accounting for the AGN component reduces the stellar mass, the galaxies are still more abundant than previously expected, a result consistent with increased star formation efficiencies implying a more shallow evolution of the volume density \citep{Somerville2025}. 

A new class of high-redshift sources discovered by JWST are the  Little Red Dot (LRD) galaxies \citep{Matthee2024, Greene2024, Akins2025, Zhang2025}, occuring predominantly at redshifts of $4-8$. These represent highly compact objects with typical diameters of less than $300$~pc and red, typically V-shaped spectra. If their luminosity is interpreted to be due to stellar luminosities, they imply  large central stellar densities  with a median of $\sim10^4$~M$_\odot$~pc$^{-3}$ and extending to maximum values of $\sim10^8$~M$_\odot$~pc$^{-3}$ \citep{Guia2024}. Because of this, these systems were proposed to be ideal environments for the formation  of supermassive black holes (SMBHs) via collision-based channels \citep{Escala2025, Pacucci2025, Dekel2025}.

The presence of high stellar densities is not restricted to LRDs or some special objects, but a more generic phenomenon in the early Universe. JWST has  detected young massive clusters (YMCs)  in several high-redshift galaxies via strong gravitational lensing. For example, \citet{Vanzella2022a} detected massive young star clusters in the strongly lensed Sunburst Lyman-continuum galaxy at $z = 2.37$, with dynamical cluster masses of order $10^7$~M$_\odot$. \citet{Vanzella2022b} found massive $10^6$~M$_\odot$ clusters in the Hubble Frontier Field A2744 at redshift $z=4$. \citet{Adamo2024} inferred bound massive clusters in the Cosmic Gems, an ultraviolet faint galaxy at $z\sim10.2$, with high stellar surface densities of $10^5$~M$_\odot$~pc$^{-2}$ \citep[see also][]{Messa2025, Vanzella2025}. At redshift $z=8.296$, \citet{Mowla2024} found a set of massive clusters cocooned in a diffuse arc termed as the Firefly Sparkle. The fact that several such systems were found, despite the requirements for strong gravitational lensing, suggests them to be not uncommon in the early Universe. The high densities and masses in these systems suggest them to be potentially relevant as formation sites of intermediate-mass black holes (IMBHs).

Particularly relevant black hole formation scenarios in this context are the collision-based scenarios \citep{Devecchi2009, Devecchhi2012, Sakurai2017, Reinoso2018, Reinoso2020, Vergara2021, Liempi2025}. \citet{Escala2021} have shown using observational data that SMBHs are present in systems where the collision timescale is shorter than the age of the system, while systems with long collision times exhibit stable nuclear star clusters without SMBHs. A systematic influence of the collision timescale on the efficiency to form a central massive object was demonstrated by \citet{Vergara2023, Vergara2024}. The efficient formation of very massive objects in systems with short collision timescales was recently demonstrated via direct N-body simulations by \citet{Vergara2023, Rantala2025, Rantalla2025b, Vergara2025a, Vergara2025b}.

Several variants of such collision-based channels exist; for example the presence of gas may further support the formation of massive objects through the interplay of collisions and accretion \citep{Boekholt2018, Tagawa2020, Aaskar2022, Schleicher2022, Schleicher2023}. Potentially important channels also include the contraction of black hole clusters in the centers of nuclear star clusters \citep{Davies2011, Lupi2014, Kroupa2020, Chassonnery2021, Gaete2024}. 

Of course, also other black hole formation channels are potentially conceivable. Intermediate-mass black holes could be remnants from the first massive stars \citep{Bromm2002, Abel2002, Yoshida2008}; or they could be the outcome of massive black hole formation via direct collapse \citep[e.g.,][]{Koushiappas2004, Bromm2003DC,Wise2008, Begelman2009, Schleicher2010,  Latif2013BH}. The relation between direct collapse and the collision-based models is not fully clear and in fact numerical simulations have shown that it is very difficult to bring all of the gas into a central massive object without fragmentation \citep[e.g.,][]{Latif2015, Latif2016}. On the other hand, gravitational torques may lead to mergers of potential fragments with the central massive object even if fragmentation happens \citep{Inayoshi2014, LatifSchleicher2015, Suazo2019}, providing a possible intermediate regime in between a ``pure'' direct collapse and a star cluster based scenario. In that sense, it is possible that direct collapse may be considered as a case of failed star cluster formation, where gravity was so efficient that a stable cluster was unable to form.

In this work, our primary aim are the massive dense clusters detected by JWST in high redshift galaxies, and their potential to form IMBHs. In particular, we aim to provide a conservative mass estimate employing standard Fokker-Planck models for the dynamical evolution of the star-clusters together with well-established results from stellar dynamics about the mergers of stars in dense environments \citep[see][]{Portegies2002}. Our methodology for this purpose is outlined in Section \ref{sec:Methodology}, while the main results are presented in Section \ref{sec:Results}. A final summary and discussion is given in Section \ref{sec:Discussion}.

\section{Methodology} \label{sec:Methodology}

In this Section, we first describe the details of the Fokker-Planck approach adopted here in Section \ref{sec:FK}. The initial conditions, and dynamical evolution of the systems are described in Section \ref{sec:InitialConditions}. The details of the black hole formation prescription are given in Section \ref{sec:blackHoleFormation}. Finally, in Section \ref{sec:Validation}, we compare our results against N-body and Monte-Carlo models.

\subsection{Fokker-Planck model} \label{sec:FK}

We use a high-accuracy finite-element method for the Fokker–Planck equation. The solver  \texttt{PhaseFlow} is publicly available  as part of the  \texttt{Agama}\footnote{\url{https://github.com/GalacticDynamics-Oxford/Agama}} library \citep{VASILIEV2017,VASILIEV2019}.

Traditionally, the one-dimensional orbit-averaged Fokker-Planck equation is expressed in terms of the energy $E$ that in the flux-conservative form  \citep[e.g., ][]{Cohn1980, Binney2009,VASILIEV2017} is written as
\begin{equation}
    \frac{\partial[ f(E,t) g(E)]}{\partial t} = -\frac{\partial \mathcal{F}(E,t) }{\partial E}, \label{eq:FPenergy}
\end{equation}
where $f(E,t)$ is the distribution function, $g(E)$ is the density of states defined as the partial derivative of the phase volume respect to the energy, and $\mathcal{F}(E, t)$ is the flux in energy space.

In \texttt{PhaseFlow}, the Fokker-Planck formalism  is reformulated and uses the phase volume $h$ (defined as the volume of phase space enclosed by the energy hypersurface) as argument of the distribution function instead of the energy $E$. 
Furthermore, it is possible to add a source term $s$ (e.g., to mimic star formation in clusters) and a sink term $\nu f$ (e.g., loss-cone draining rate) that results in the following expression:
\begin{equation}
    \frac{\partial f(h,t)}{\partial t} = -\frac{\partial \mathcal{F}(h,t)}{\partial h} +s(h,t)-\nu(h,t) f(h,t),
\end{equation}
with the distribution function $f(h,t)$ and the flux $\mathcal{F}(h,t)$ both now as function of the phase volume. We here adopt $s(h,t)=0$ and $\nu(h,t)=0$, as we neglect star formation and we assume that at least initially an IMBH is not yet present (thus no loss cone effects). The flux through the phase volume is given as \begin{equation}
-\mathcal{F}(h,t)=A(h)f(h,t)+D(h)\frac{\partial f(h,t)}{\partial h},
\end{equation}
with $A(h)$ and $D(h)$ the advection and diffusion coefficients, respectively. 

In our model, we adopt a logarithmically spaced phase volume grid that contains $200$ points. The minimum value ($h_{\rm min}$) is set equals to $10^{-20}$, while the maximum value adopted is $h_{\rm max}=10^3$,  In our adopted virial units $(G=M=1)$, the total bound phase volume of the cluster is of order unity. Therefore, $h_{\rm max}=10^3$ safely over-encompasses the bound phase space, ensuring accurate energy diffusion for loosely bound stars approaching the escape energy without boundary truncation. Conversely, because the core phase volume scales roughly as $h_{\rm core} \propto M_c^{3/2} r_c^{3/2}$, it drops exponentially during deep core collapse. The extreme lower bound of $h_{\rm min}=10^-20$ is mathematically required to maintain flux conservation and continuously resolve the distribution function as the core radius $r_c$ shrinks and the central density diverges. We use the \cite{CHANG1970} discretization scheme with a timestep adaptively   set with an accuracy parameter $\epsilon=10^{-4}$ to ensure flux conservation.

\subsection{Initial conditions and dynamical evolution of the system}\label{sec:InitialConditions}

The Fokker-Planck model implicitly assumes that all our clusters are composed of equal-mass stars spatially distributed following a Plummer density profile \citep{PLUMMER1911}, 
\begin{equation}
    \rho(r)= \frac{3M}{4\pi b^3}\left(1+\frac{r^2}{b^2}\right)^{-\frac{5}{2}},
\end{equation}
where $M$ is the total mass of the system and $b$ is the Plummer radius. For the simulations, we adopt virial units ($G=M=1$) with a total energy of $E_{\rm tot} = -1/4$. This scaling constrains the Plummer scale radius to $b \approx 0.589$ (specifically $b = 3\pi/16$ in standard virial units). Assuming equal-mass stars is clearly a simplified assumption, as realistic star clusters may rather evolve on the mass segregation timescale which is shorter than the relaxation time. Considering mass segregation effects further would favour the subsequent formation of a central massive object. The Fokker-Planck model adopted here will therefore allow us to obtain a conservative estimate, while the evolution in real star clusters is potentially accelerated. The relaxation time of the system is given as
\begin{equation}
    t_{\rm relax} = \frac{0.206 M^{1/2} b^{3/2}}{G^{1/2} m_{\star} \ln \Lambda},\label{relax}
\end{equation}
where $m_{\star}$ represents the individual stellar mass and $\ln \Lambda = \ln{\gamma N}$ is the Coulomb logarithm,  where $\gamma=0.4$ is adopted for systems composed by equal-mass stars \citep{Spitzer1987}. In consequence the number of stars $N$ is given by $N=M_{\rm cl}/ 1\,{\rm M}_\odot$ (i.e., assuming solar-mass stars) with $M_{\rm cl}$ the stellar mass of the cluster in solar masses.

As the YMCs observed with JWST have typical ages of a few ten million years \citep{Vanzella2022a, Vanzella2022b, Vanzella2023, Adamo2024, Mowla2024}, we here evolve the Fokker-Planck model for a characteristic timescale of about $100$~Myr, to account for the fact that the evolution in realistic star clusters will be faster due to mass segregation effects compared to models assuming equal mass stars. Indeed, equal-mass systems typically reaches core collapse on a timescale of $\sim 15 t_{\rm relax}$ \citep{Cohn1980}, while the presence of a stellar mass function can shorten this process to as little as $2-3\,t_{\rm relax}$ \citep[e.g.,][]{Gurkan2004} as massive stars rapidly sink to the center. In real systems, there is even the possibility of pre-existing central cores resulting from the process of star cluster formation, while our model conservatively starts from a Plummer sphere for the initial distribution of stars.

\begin{table}
\caption{\label{tab:cluster_properties}Properties of the high-redshift massive star clusters.}
\centering
\begin{tabular}{lccc}
\hline\hline
Cluster ID & $M_{\rm cl}$ [$M_\odot$] & $R_h$ [pc] & $Z$ [$Z_\odot$] \\
\hline
A1\tablefootmark{a}    & $2.45 \times 10^{6}$ & $1.1$ & 0.005 \\
B1\tablefootmark{a}    & $2.65 \times 10^{6}$ & $1.1$ & 0.005 \\
C1\tablefootmark{a}    & $1.13 \times 10^{6}$ & $<1$  & 0.005 \\
D1\tablefootmark{a}    & $1.13 \times 10^{6}$ & $0.6$ & 0.005 \\
E1\tablefootmark{a}    & $1.01 \times 10^{6}$ & $0.4$ & 0.005 \\
\hline
FF-1\tablefootmark{b}  & $1.82 \times 10^{5}$ & $<6.8$ & 0.02 \\
FF-2\tablefootmark{b}  & $2.29 \times 10^{6}$ & $<6.2$ & 0.02 \\
FF-3\tablefootmark{b}  & $4.57 \times 10^{5}$ & $<5.1$ & 0.02 \\
FF-4\tablefootmark{b}  & $1.38 \times 10^{5}$ & $<4.9$ & 0.02 \\
FF-5\tablefootmark{b}  & $2.57 \times 10^{5}$ & $<4.6$ & 0.02 \\
FF-6\tablefootmark{b}  & $2.04 \times 10^{5}$ & $<4.3$ & 0.02 \\
FF-7\tablefootmark{b}  & $1.42 \times 10^{5}$ & $<4.1$ & 0.02 \\
FF-8\tablefootmark{b}  & $1.02 \times 10^{5}$ & $<3.9$ & 0.02 \\
FF-9\tablefootmark{b}  & $9.33 \times 10^{4}$ & $<4.2$ & 0.02 \\
FF-10\tablefootmark{b} & $1.00 \times 10^{6}$ & $<4.3$ & 0.02 \\
\hline
5.1a\tablefootmark{c}  & $9.10 \times 10^{6}$ & $8.1$  & 0.02 \\
5.1b\tablefootmark{c}  & $1.02 \times 10^{7}$ & $8.2$  & 0.02 \\
5.1c\tablefootmark{c}  & $1.13 \times 10^{7}$ & $9.6$  & 0.02 \\
5.1h\tablefootmark{c}  & $1.22 \times 10^{7}$ & $10.8$ & 0.02 \\
5.1i\tablefootmark{c}  & $5.10 \times 10^{6}$ & $7.8$  & 0.02 \\
5.11\tablefootmark{c}  & $1.17 \times 10^{7}$ & $8.5$  & 0.02 \\
5.1m\tablefootmark{c}  & $1.28 \times 10^{7}$ & $<19.5$& 0.02 \\
5.1n\tablefootmark{c}  & $1.71 \times 10^{7}$ & $<20.4$& 0.02 \\
5.2h\tablefootmark{c}  & $5.40 \times 10^{6}$ & $4.8$  & 0.02 \\
5.3h\tablefootmark{c}  & $1.47 \times 10^{7}$ & $23.7$ & 0.02 \\
5.4a\tablefootmark{c}  & $3.20 \times 10^{6}$ & $5.4$  & 0.02 \\
5.5a\tablefootmark{c}  & $3.00 \times 10^{6}$ & $7.9$  & 0.02 \\
5.6a\tablefootmark{c}  & $3.20 \times 10^{6}$ & $9.0$  & 0.02 \\
5.8d\tablefootmark{c}  & $7.10 \times 10^{6}$ & $20.2$ & 0.02 \\
5.9d\tablefootmark{c}  & $4.50 \times 10^{6}$ & $21.4$ & 0.02 \\
5.11d\tablefootmark{c} & $2.40 \times 10^{6}$ & $15.0$ & 0.02 \\
5.16d\tablefootmark{c} & $5.50 \times 10^{6}$ & $22.0$ & 0.02 \\
5.12g\tablefootmark{c} & $1.00 \times 10^{6}$ & $2.9$  & 0.02 \\
5.13g\tablefootmark{c} & $1.00 \times 10^{5}$ & $0.9$  & 0.02 \\
5.15h\tablefootmark{c} & $1.00 \times 10^{5}$ & $1.3$  & 0.02 \\
\hline
3.1a\tablefootmark{d}  & $4.10 \times 10^{6}$ & $13.8$ & 0.02 \\
3.2a\tablefootmark{d}  & $7.00 \times 10^{5}$ & $3.2$  & 0.02 \\
3.3a\tablefootmark{d}  & $9.00 \times 10^{5}$ & $11.4$ & 0.02 \\
3.1b\tablefootmark{d}  & $3.60 \times 10^{6}$ & $15.3$ & 0.02 \\
3.2b\tablefootmark{d}  & $5.00 \times 10^{5}$ & $3.0$  & 0.02 \\
3.3b\tablefootmark{d}  & $1.20 \times 10^{6}$ & $10.9$ & 0.02 \\
3c\tablefootmark{d}    & $1.65 \times 10^{8}$ & $279.4$& 0.02 \\
\hline
1b\tablefootmark{e}    & $7.10 \times 10^{6}$ & $1.4$  & 0.02 \\
2b\tablefootmark{e}    & $3.90 \times 10^{6}$ & $6.3$  & 0.02 \\
3b\tablefootmark{e}    & $1.10 \times 10^{6}$ & $6.1$  & 0.02 \\
4b\tablefootmark{e}    & $1.01 \times 10^{7}$ & $24.8$ & 0.02 \\
5b\tablefootmark{e}    & $3.10 \times 10^{6}$ & $4.9$  & 0.02 \\
6b\tablefootmark{e}    & $3.30 \times 10^{6}$ & $8.5$  & 0.02 \\
\hline
\end{tabular}
\tablefoot{
\tablefoottext{a}{\cite{Adamo2024}}
\tablefoottext{b}{\cite{Mowla2024}}
\tablefoottext{c}{\cite{Vanzella2022a}}
\tablefoottext{d}{\cite{Vanzella2022b}}
\tablefoottext{e}{\cite{Vanzella2023}}
}
\end{table}
\subsection{Black hole mass estimation}\label{sec:blackHoleFormation}

The contraction of the star cluster as described through the Fokker-Planck model will lead to the formation of a central core within the star cluster. The growth of central massive objects in such environments via runaway collisions has been studied by \citet[][]{Portegies2002, Katz2015, Rantalla2025b, Vergara2025a, Vergara2025b}, where a detailed analytical framework has been provided by \citet{Portegies2002} considering the migration times of stars of different masses. This model has been recently extended by \citet{Fujii2024, Pacucci2025} to account for the mass loss through winds by the very massive star (VMS).  The maximum mass (assuming that all the collisions involve the same star) is given by 
\begin{equation}
    \dot{M}_{\rm acc} = \dot{N}_{\rm coll}\langle \delta m \rangle_{\rm coll}, \label{eq:massRateVMS}
\end{equation}
with $\dot{N}_{\rm coll}$ the average collision rate and $\langle m \rangle_{\rm coll}$ the average mass increase per collision. 

In \citet{Portegies2002}, their suite of simulations showed that collisions  between stars generally occur in dynamically formed (‘‘three body’’) binaries and, in consequence, the average collision rate $\dot{N}_{\rm coll}$ is related to the binary formation rate. Neglecting stellar evolution effects and assuming the  large-scale energy flux in the cluster to be powered by binary heating in the core, the average collision rate is well approximated by
\begin{equation}
    \dot{N}_{\rm coll} \approx 10^{-3} f_c \frac{N}{t_{\rm 
    relax}}, \label{eq:collRate}
\end{equation}
where $f_c\leq 1$ represents the effective fraction of dynamically formed binaries that produce a collision, $t_{\rm  relax}$ is the relaxation time, and $N$ the number of stars in the system, estimated as $M_{\rm cl}/1\, {\rm M}_\odot$. We here employ the value of $f_c=0.8$ as found empirically by \citet{Fujii2024} and later adopted in the work of  \citet{Pacucci2025}. While the negligence of stellar evolution effects is clearly a simplifying assumption, we will in the following verify the results obtained with this model through the comparison with direct N-body simulations including stellar evolution to show that the simplification still leads to reasonable results.

The average mass increase per collision is given by 
\begin{equation}
    \langle \delta m\rangle_{\rm coll}  \simeq 4\frac{t_{\rm relax}}{t}\langle m \rangle \ln{\Lambda},\label{eq:massRate}
\end{equation}
where $\langle m \rangle = 1\,{\rm M}_\odot$, and again $\ln{\Lambda}=\ln{0.4N}$. Thus, replacing Eqs. \ref{eq:collRate} and \ref{eq:massRate} in Eq. \ref{eq:massRateVMS}, the mass growth of the VMS due to stellar collisions is
\begin{equation}
\dot{M}_{\rm acc} = 4\times 10^{-3} f_c \ln{\Lambda}\frac{M_{\rm cl}}{t}. \label{eq:massVMS2}
\end{equation}

Following the procedure of \cite{Fujii2024}, Eq. \ref{eq:massVMS2} can be rewritten defining a  "supply rate" $\dot{M}_{\rm cl} = M_{\rm cl}/t$. However, we must account for the fact that the runaway collision process is dynamically restricted to the dense central region of the cluster. While the global energy budget is determined by the total cluster mass, the immediate mass reservoir available for the growth of the VMS is the cluster core.   Direct N-body simulations for example by \citet{Arca2023,   Vergara2025a, Rantala2026} show that the timescale over which stars are supplied to the central object is comparable to the relaxation timescale of the central core, so that we estimate $t$ evaluating the relaxation time (Eq.~\ref{relax}) using the core mass $M_{\rm core}$, which we also employ for estimating the stellar mass reservoir.  We identify the core radius ($r_{\rm core}$) as the point where the density profile drops to half its central value, and define $M_{\rm core}$ as the enclosed mass at this radius. The effective  supply rate feeding the VMS is therefore defined as ${\rm SF} = M_{\rm core}/t_{\rm relax,core}$,  where $t_{\rm relax,core}$ is the relaxation time of the core. 

Substituting $M_{\rm cl}$ with $M_{\rm core}$ in Eq. \ref{eq:massVMS2}, the "accretion" rate becomes dependent on the Coulomb logarithm $\ln \Lambda$. For the massive clusters considered in this work ($N \sim 10^5 - 10^6$), the Coulomb logarithm is in the range $\ln \Lambda \approx 10-12$. Thus, the pre-factor $0.003 \ln \Lambda$ is well-approximated by a constant efficiency of $\approx 0.03$. This yields the final expression used in our model:

\begin{equation}
    \dot{M}_{\rm acc} \approx 0.03 \times \frac{M_{\rm core}}{t_{\rm relax,core}}. \label{eq:finalMassAcc}
\end{equation}

This "supply" rate is then balanced against the wind mass loss rate. The mass loss rate (assuming fixed metallicity) is given by \citet{VINK2018} 
\begin{equation}
\log_{10}\left(\frac{\dot{M}_{\rm wind}}{{\rm M_{\odot}\,yr^{-1}}}\right)=-9.13+2.1\log_{10}{\left(\frac{M_{\rm VMS}}{{\rm M_{\odot}}}\right)}+0.74\log_{10}{\left(\frac{Z}{{\rm Z_\odot}}\right)}. \label{eq:massLossRate}   
\end{equation}

The time evolution of the mass of the VMS is given by the balance between mass gain via collisions and mass loss due to stellar winds:
\begin{equation}
    \dot{M}_{\rm VMS} = \dot{M}_{\rm acc}-\dot{M}_{\rm wind}.\label{massev}
\end{equation}

We assume here that the VMS grows until the mass loss from the wind becomes comparable to the mass growth via collisions.  Some uncertainty in this approach lies in the timescale that is adopted here for the timescale entering in Eq.~\ref{eq:massVMS2}, which could vary for example in the presence of rotation or depending on the binary fraction. However, given the balance in Eq.~\ref{massev} between mass gain by accretion and mass loss by wind, the dependence of the final mass of the VMS on the timescale is relatively weak and would scale as $t^{1/2.1}\sim t^{0.48}$.  A change in the timescale by a factor of $10$ thus will not affect the result by more than a factor of $3$, consistent with the level of uncertainty in the absence of detailed N-body simulations. 

To estimate the final BH seed mass ($M_{\rm BH}$), we assume that the VMS collapses directly into aBH at the end of its life. Given the high masses ($> 10^3 \, {\rm M}_\odot$) and low metallicities considered in this work, we assume negligible mass loss during the collapse phase, setting $M_{\rm BH} \approx M_{\rm VMS}$. 

\subsection{Comparison with numerical simulations} \label{sec:Validation}

Many of the star clusters found by JWST are quite massive, making direct N-body simulations very difficult or essentially unfeasible. We thus necessarily need to employ a simplified framework in order to derive estimates for the masses of possible IMBHs. Nonetheless, it is instructive to test our model framework in the regime where such a comparison is possible. For this purpose, we utilize  large N-body simulations by \citet{Arca2023},  \citet{Vergara2025a}  based on the \textsc{nbody6++gpu} and \textsc{MOCCA} codes, as well as the FROST clusters presented by and \citet{Rantala2026}. Both approaches incorporate up-to-date stellar evolution routines (SSE/BSE) alongside specific prescriptions for the formation and dynamical evolution of VMSs.

\begin{figure}
    \centering
    \includegraphics[width=\columnwidth]{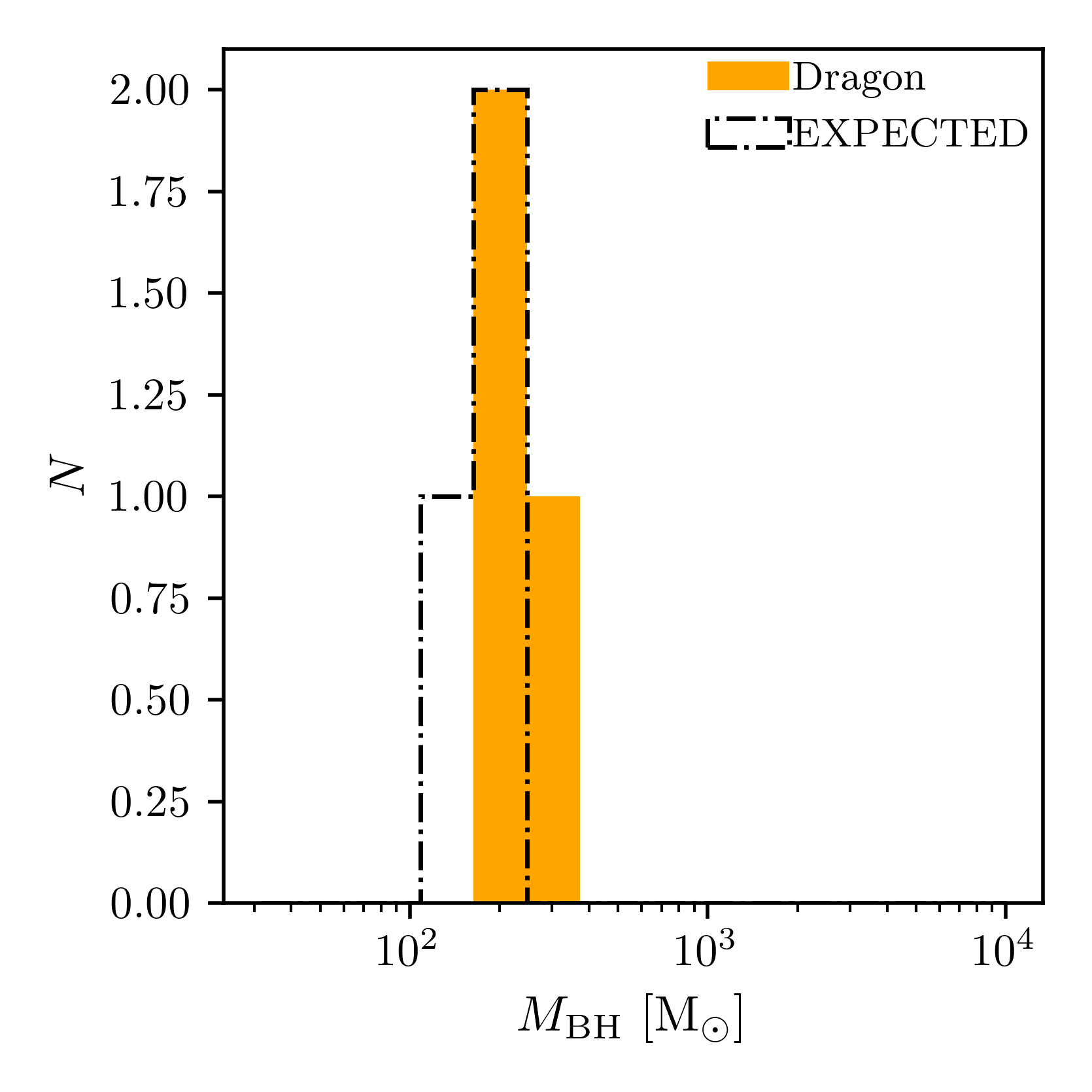}
    \caption{ Comparison of the black hole mass distribution. The black dashed line shows the prediction from the framework presented here, while the orange histogram represents the results from the Dragon simulations \citep{Arca2023}.
    }
    \label{fig:DragonComparison}
\end{figure}

\begin{figure}
    \centering
    \includegraphics[width=\columnwidth]{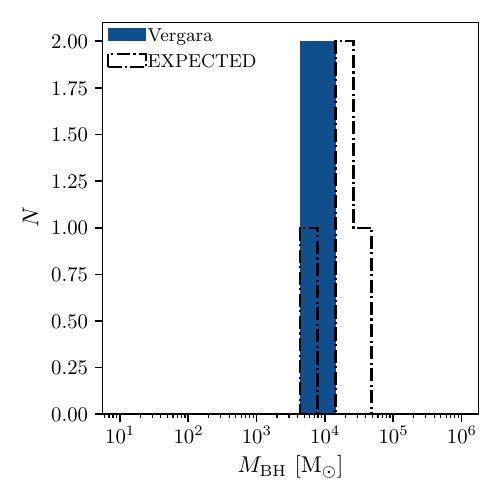}
    \caption{Comparison of the black hole mass distribution. The black dashed line shows the prediction from the framework presented here, while the blue histogram represents the results from the N-body simulations of \citet{Vergara2025a}. }
    \label{fig:VergaraComparison}
\end{figure}

\begin{figure}
    \centering
    \includegraphics[width=\columnwidth]{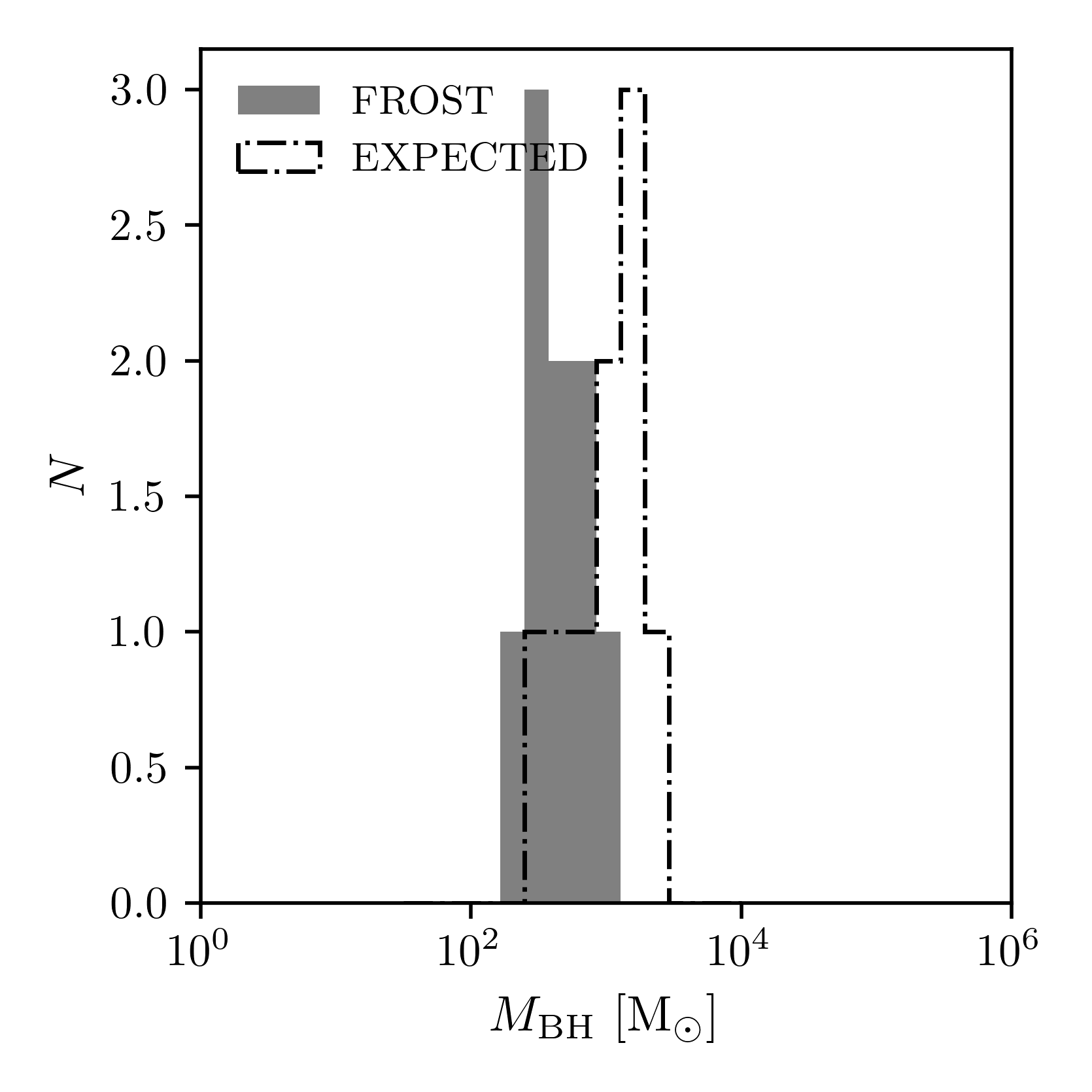}
    \caption{ Comparison of the black hole mass distribution. The black dashed line shows the prediction from the framework presented here, while the grey histogram represents the results from the FROST simulations \citep{Rantala2026}.}
    \label{fig:FROSTComparison}
\end{figure}

To ensure a consistent comparison, we initialized our semi-analytic model using the cluster parameters provided in the original papers.  In Fig.~\ref{fig:DragonComparison}, we compare our results to the DRAGON~II simulations \citep{Arca2023}, which predict IMBH masses of $\sim150-350$~M$_\odot$. These are slightly underestimated in our model framework, which predicts typical values of $100-250$~M$_\odot$.  Figure \ref{fig:VergaraComparison} illustrates the results of this cross-check with the \citet{Vergara2025a} simulations. The blue histogram shows the distribution of black hole masses obtained from the \citet{Vergara2025b} simulations, while the black dot-dashed step line shows the mass distribution predicted by our analytic model for the same set of clusters. Our model framework predicts masses of $8\times10^3 - 5\times10^4$~M$_\odot$ for the \citet{Vergara2025b} simulations, while the simulations themselves derived masses in the range of $4\times10^3-1.5\times10^4$~M$_\odot$. In this case, the model captures the magnitude of the expected masses but with deviations including a factor of $3-5$.

 In Fig.~\ref{fig:FROSTComparison}, we compare with the FROST clusters by \citet{Rantala2026}, where the original simulations predicted masses of $\sim200-1000$~M$_\odot$, while our framework suggests masses of $300-3000$~M$_\odot$. In this case our framework shows an uncertainty within a factor of $2-3$. In summary, we can say that our model framework  provides the order-of-magnitude of the massive objects formed in these simulations, with the expected deviations in the range of  around half an order of magnitude when compared to detailed N-body simulations. While our model does not include an explicit treatment of the mass loss during collisions, we note that such a treatment is included in the simulations of \citet{Vergara2025a} and \citet{Rantala2026} following \citet{Glebbeek2008, Glebbeek2009, Glebbeek2013} and thus consistent with the overall uncertainty considered here.

\section{Results} \label{sec:Results}

In the following subsections, we employ the framework presented above to the different systems for which JWST has provided measurements of the masses and radii of young massive clusters. We also aim to summarize global scaling relations and efficiencies.

\subsection{Fokker-Planck results}

\begin{figure}[h!]
    \centering
    \includegraphics{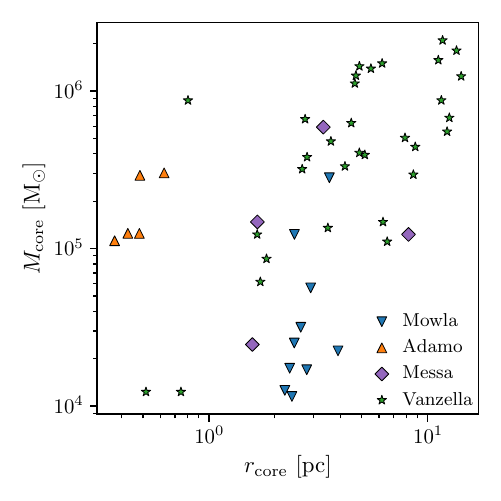}
    \caption{Mass of the core as a function of the core radius. Blue triangles are clusters from \citet{Mowla2024}, orange triangles clusters from \citet{Adamo2024}, purple diamonds data from \citet{Messa2025}, and green stars show data from \citet[][]{Vanzella2022a,Vanzella2022b,Vanzella2023}.}
    \label{fig:global}
\end{figure}

The results from the Fokker-Planck model indicate the evolution of the star clusters we can expect after a timescale of $100$~Myr, assuming clusters of equal mass stars. In real stellar clusters, as discussed above, the evolution can be easily accelerated at least by a factor of a few, considering that the mass segregation timescale is shorter than the relaxation time as well as even the presence of a possible primordial mass segregation. The expected properties of the resulting cores are given in Fig.~\ref{fig:global} showing core mass ($M_{\rm core}$) as a function of core radius ($r_{\rm core}$). The core masses tend to be in the range from $10^4$~M$_\odot$ up to $10^6$~M$_\odot$, with radii in the range from $\sim0.4$~pc up to $\sim15$~pc. We note that the cores from the \citet{Adamo2024} sample are particularly compact of the order $\sim0.5$~pc, and also the \citet{Vanzella2023} sample includes some clusters with core radii of $\sim0.6-0.8$~pc, even if most of their cores have radii of $\sim1.5-15$~pc. The data from \citet{Messa2025} and \citet{Mowla2024} share the mass range but tend to be at the larger radii. The trends of the cores follow the underlying properties of the star clusters. Globally, this is consistent both with expectations from numerical simulations, which show that about $10\%$ of the star clusters might be on the rather compact side \citep{Grudic2023}, as well as with the properties of star clusters in the local Universe, where again about $10\%$ show rather compact radii compatible with the JWST clusters \citep{Brown2021}. The latter is compatible also with the \citet{Marks2012} relation, indicating compact cluster radii at the initial formation time of the clusters.

\subsection{Compact clusters in the Cosmic Gems Arc}

We now use the framework laid out in section~2.3 to estimate the masses of IMBHs that could form in the dense JWST clusters in a collision-based scenario. We emphasize that these estimates are subject to uncertainties of at least a factor of a few, as most of the clusters are in a regime where direct N-body simulations are not feasible. Nonetheless deriving such estimates based on the previous results of \citet{Portegies2002} and the framework of \citet{Pacucci2025} will be useful to assess the possible mass range of IMBHs that could be present.

\begin{table}[h!]
    \centering
    \caption{Estimated core and black hole properties for the Cosmic Gems cluster sample.}
    \label{tab:adamo_results}
    \begin{tabular}{l c c c c c}
    \hline \hline
    ID & $M_{\rm cl}$ & $R_h$ & $M_{\rm core}$ & $R_{\rm core}$ & $M_{\rm BH}$ \\
     & $[{\rm M}_\odot]$ & $[{\rm pc}]$ & $[{\rm M}_\odot]$ & $[{\rm pc}]$ & $[{\rm M}_\odot]$ \\
    \hline
    A1 & $2.45 \times 10^6$ & 1.10 & $3.03 \times 10^5$ & 0.62 & $2.68 \times 10^3$ \\
    B1 & $2.65 \times 10^6$ & 0.90 & $2.92 \times 10^5$ & 0.48 & $2.64 \times 10^3$ \\
    C1 & $1.13 \times 10^6$ & 0.90 & $1.25 \times 10^5$ & 0.48 & $1.76 \times 10^3$ \\
    D1 & $1.13 \times 10^6$ & 0.80 & $1.25 \times 10^5$ & 0.43 & $1.76 \times 10^3$ \\
    E1 & $1.01 \times 10^6$ & 0.70 & $1.12 \times 10^5$ & 0.37 & $1.67 \times 10^3$ \\
    \hline
    \end{tabular}
    \tablefoot{
Properties are derived from the cluster data presented in \cite{Adamo2024}.
}
\end{table}

 The results for the clusters identified in the Cosmic Gems arc are presented in Table \ref{tab:adamo_results}. These systems are characterized by their extreme compactness, with half-mass radii consistently around $R_h \approx 1$ pc despite having stellar masses in the range of $10^6 \, {\rm M}_\odot$. The high stellar density facilitates a rapid core collapse, leading to substantial core masses ($M_{\rm core} \sim 1-3 \times 10^5 \, {\rm M}_\odot$). Consequently, our model predicts the formation of IMBHs with masses  ranging from $\sim1.7\times10^3$ to $\sim 2.7\times 10^3 \, {\rm M}_\odot$. The low metallicity assumed for this high-redshift galaxy ($Z \approx 0.005 \, {\rm Z}_\odot$) further aids in retaining the accreted mass by reducing the efficiency of stellar wind mass loss during the VMS phase.

\begin{figure}[h!]
    \centering
    \includegraphics{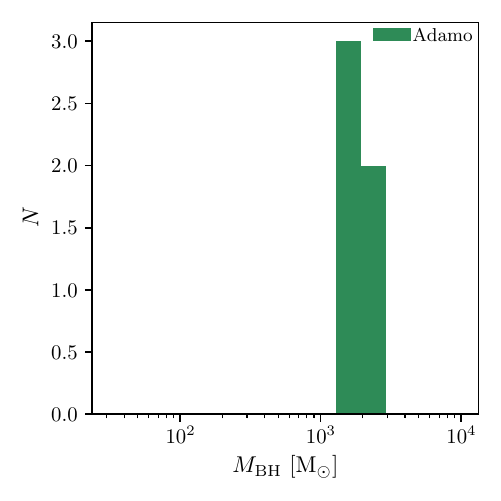}
    \caption{Mass distribution of estimated black hole masses in the data sample of \cite{Adamo2024}.}
    \label{fig:adamoDistribution}
\end{figure}

 The results of the calculation are provided in Fig.~\ref{fig:adamoDistribution}. As the star clusters and cores have rather similar properties, the expected  mass distribution of the IMBHs shows a clear peak around  $\sim10^3$~M$_\odot$, providing a significant potential to form quite massive black holes.

\subsection{Extended clusters in the Firefly Sparkle}

\begin{figure}[h!]
    \centering
    \includegraphics{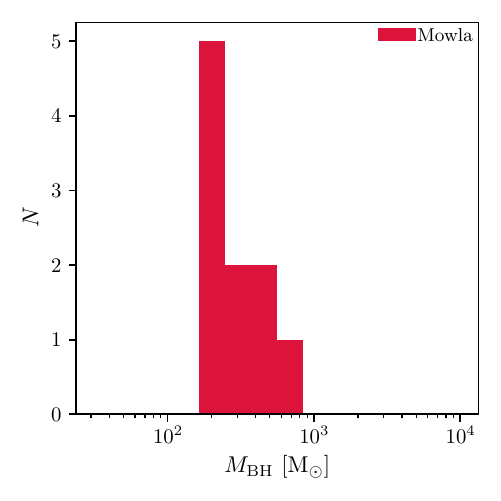}
    \caption{Mass distribution  of estimated black hole masses for the \citet{Mowla2024} sample.}
    \label{fig:mowlaDistribution}
\end{figure}

Table \ref{tab:mowla_results} summarizes the properties of the clusters associated with the Firefly Sparkle \citep{Mowla2024}. In contrast to the Cosmic Gems, these clusters exhibit significantly larger radii ($R_h \sim 4-6$ pc) for typically lower total masses ($10^5 - 10^6 \, {\rm M}_\odot$). This lower dynamical density results in less massive cores ($M_{\rm core} \sim 10^4-3 \times 10^5 \, {\rm M}_\odot$) (see Fig.~\ref{fig:global}), limiting the reservoir available for the runaway collision process. As a result, the predicted black hole masses are also reduced, with the distribution (Fig.~\ref{fig:mowlaDistribution}) in the range of $1.8\times10^2 -8.1\times10^2 \, {\rm M}_\odot$. Within the sources investigated here, these objects are forming the lower boundary of expected IMBH masses. 

\begin{table}[h!]
    \centering
\caption{Estimated core and black hole properties for the Firefly Sparkle cluster sample.}
\label{tab:mowla_results}
    \begin{tabular}{l c c c c c}
    \hline \hline
    ID & $M_{\rm cl}$ & $R_h$ & $M_{\rm core}$ & $R_{\rm core}$ & $M_{\rm BH}$ \\
     & $[{\rm M}_\odot]$ & $[{\rm pc}]$ & $[{\rm M}_\odot]$ & $[{\rm pc}]$ & $[{\rm M}_\odot]$ \\
    \hline
    FF-1 & $1.82 \times 10^5$ & 6.80 & $2.24 \times 10^4$ & 3.89 & $2.42 \times 10^2$ \\
    FF-2 & $2.29 \times 10^6$ & 6.20 & $2.82 \times 10^5$ & 3.55 & $8.07 \times 10^2$ \\
    FF-3 & $4.57 \times 10^5$ & 5.10 & $5.63 \times 10^4$ & 2.92 & $3.74 \times 10^2$ \\
    FF-4 & $1.38 \times 10^5$ & 4.90 & $1.70 \times 10^4$ & 2.80 & $2.12 \times 10^2$ \\
    FF-5 & $2.57 \times 10^5$ & 4.60 & $3.17 \times 10^4$ & 2.63 & $2.85 \times 10^2$ \\
    FF-6 & $2.04 \times 10^5$ & 4.30 & $2.52 \times 10^4$ & 2.46 & $2.55 \times 10^2$ \\
    FF-7 & $1.41 \times 10^5$ & 4.10 & $1.74 \times 10^4$ & 2.34 & $2.14 \times 10^2$ \\
    FF-8 & $1.02 \times 10^5$ & 3.90 & $1.26 \times 10^4$ & 2.23 & $1.83 \times 10^2$ \\
    FF-9 & $9.33 \times 10^4$ & 4.20 & $1.15 \times 10^4$ & 2.40 & $1.76 \times 10^2$ \\
    FF-10 & $1.00 \times 10^6$ & 4.30 & $1.23 \times 10^5$ & 2.46 & $5.44 \times 10^2$ \\
    \hline
    \end{tabular}
    \tablefoot{
Properties are derived from the cluster data presented in \cite{Mowla2024}.
}
\end{table}

\subsection{Variable density candidates}

\begin{table}[h!]
    \centering
    \caption{Core and black hole properties for the Variable density candidates.}
    \label{tab:messa_results}
    \begin{tabular}{l c c c c c}
    \hline \hline
    ID & $M_{\rm cl}$ & $R_h$ & $M_{\rm core}$ & $R_{\rm core}$ & $M_{\rm BH}$ \\
     & $[{\rm M}_\odot]$ & $[{\rm pc}]$ & $[{\rm M}_\odot]$ & $[{\rm pc}]$ & $[{\rm M}_\odot]$ \\
    \hline
    M1 & $4.80 \times 10^6$ & 5.82 & $5.91 \times 10^5$ & 3.34 & $3.69 \times 10^3$ \\
    M2 & $1.00 \times 10^6$ & 14.25 & $1.23 \times 10^5$ & 8.18 & $1.75 \times 10^3$ \\
    M3 & $2.00 \times 10^5$ & 2.76 & $2.46 \times 10^4$ & 1.58 & $8.13 \times 10^2$ \\
    M4 & $1.20 \times 10^6$ & 2.91 & $1.48 \times 10^5$ & 1.67 & $1.91 \times 10^3$ \\
    \hline
    \end{tabular}
    \tablefoot{
Properties are derived from the cluster data presented in \cite{Messa2025}.
}
\end{table}

The results for the cluster candidates identified by \citet{Messa2025} are listed in Table \ref{tab:messa_results}. This sample presents a diverse set of environments, ranging from compact ($R_h \sim 2.7$ pc) to extended ($R_h \sim 14$ pc) systems. Notably, the most massive cluster in this set (M1, $M_{\rm cl} \approx 4.8 \times 10^6 \, {\rm M}_\odot$) yields a  black hole mass of $\sim3.7 \times 10^3 \, {\rm M}_\odot$. Despite the high total cluster mass, the BH formation efficiency is moderated by the higher assumed metallicity ($Z=0.1 \, {\rm Z}_\odot$), implying more mass loss through winds, and the relatively larger radii compared to the \citet{Adamo2024} sample. 

\begin{figure}[h!]
    \centering
    \includegraphics{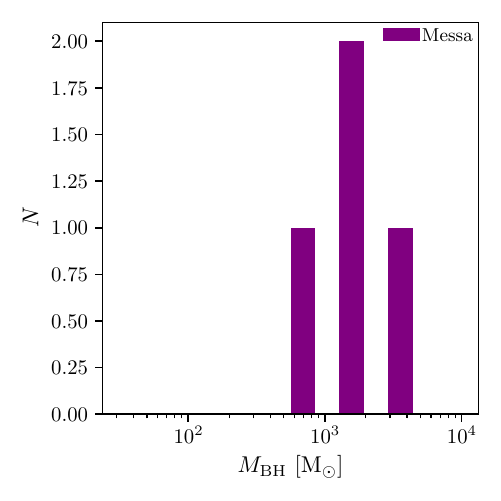}
    \caption{Distribution of estimated black hole masses for the \citet{Messa2025} clusters. }
    \label{fig:messaDistribution}
\end{figure}

The expected distribution of the IMBH masses is given in Fig.~\ref{fig:messaDistribution}, with  masses ranging from $\sim 8.1\times 10^2 \,{\rm M}_{\odot}$ to $\sim 3.7\times10^3\, {\rm M}_{\odot}$.

\subsection{Massive young clusters in the Sunburst Arc and Frontier Fields}

\begin{figure}[h!]
    \centering
    \includegraphics{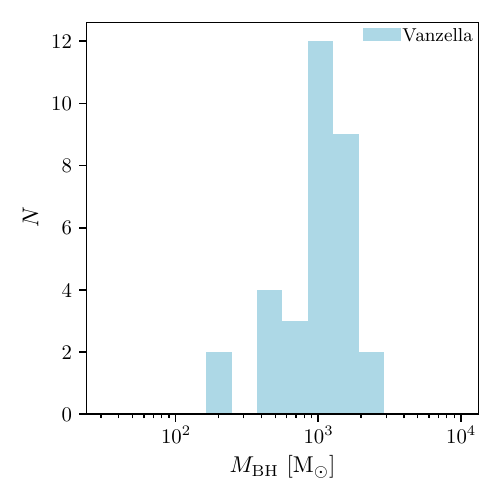}
    \caption{Mass distribution of estimated black hole masses for the \citet{Vanzella2022a,Vanzella2022b,Vanzella2023} samples.}
    \label{fig:vanzelladistribution}
\end{figure}

\begin{table}[h!]
    \centering
    \caption{Estimated core and black hole properties for the Sunburst Arc and Frontier Fields cluster samples.}
    \label{tab:vanzella_results}
    \begin{tabular}{l c c c c c}
    \hline \hline
    ID & $M_{\rm cl}$ & $R_h$ & $M_{\rm core}$ & $R_{\rm core}$ & $M_{\rm BH}$ \\
     & $[{\rm M}_\odot]$ & $[{\rm pc}]$ & $[{\rm M}_\odot]$ & $[{\rm pc}]$ & $[{\rm M}_\odot]$ \\
    \hline
    5.1a & $9.10 \times 10^6$ & 8.10 & $1.12 \times 10^6$ & 4.65 & $1.56 \times 10^3$ \\
    5.1b & $1.02 \times 10^7$ & 8.20 & $1.26 \times 10^6$ & 4.70 & $1.64 \times 10^3$ \\
    5.1c & $1.13 \times 10^7$ & 9.60 & $1.39 \times 10^6$ & 5.51 & $1.72 \times 10^3$ \\
    5.1h & $1.22 \times 10^7$ & 10.80 & $1.50 \times 10^6$ & 6.19 & $1.79 \times 10^3$ \\
    5.1i & $5.10 \times 10^6$ & 7.80 & $6.28 \times 10^5$ & 4.48 & $1.18 \times 10^3$ \\
    5.1l & $1.17 \times 10^7$ & 8.50 & $1.44 \times 10^6$ & 4.88 & $1.75 \times 10^3$ \\
    5.1m & $1.28 \times 10^7$ & 19.50 & $1.58 \times 10^6$ & 11.19 & $1.83 \times 10^3$ \\
    5.1n & $1.71 \times 10^7$ & 20.40 & $2.11 \times 10^6$ & 11.70 & $2.10 \times 10^3$ \\
    5.2h & $5.40 \times 10^6$ & 4.80 & $6.65 \times 10^5$ & 2.75 & $1.21 \times 10^3$ \\
    5.3h & $1.47 \times 10^7$ & 23.70 & $1.81 \times 10^6$ & 13.55 & $1.96 \times 10^3$ \\
    5.4a & $3.20 \times 10^6$ & 5.40 & $3.20 \times 10^5$ & 2.67 & $8.57 \times 10^2$ \\
    5.5a & $3.00 \times 10^6$ & 7.90 & $3.33 \times 10^5$ & 4.19 & $8.73 \times 10^2$ \\
    5.6a & $3.20 \times 10^6$ & 9.00 & $3.94 \times 10^5$ & 5.16 & $9.46 \times 10^2$ \\
    5.8d & $7.10 \times 10^6$ & 20.20 & $8.75 \times 10^5$ & 11.55 & $1.38 \times 10^3$ \\
    5.9d & $4.50 \times 10^6$ & 21.40 & $5.54 \times 10^5$ & 12.27 & $1.11 \times 10^3$ \\
    5.11d & $2.40 \times 10^6$ & 15.00 & $2.95 \times 10^5$ & 8.61 & $8.25 \times 10^2$ \\
    5.16d & $5.50 \times 10^6$ & 22.00 & $6.78 \times 10^5$ & 12.57 & $1.23 \times 10^3$ \\
    5.12g & $1.00 \times 10^6$ & 2.90 & $1.23 \times 10^5$ & 1.66 & $5.44 \times 10^2$ \\
    5.13g & $1.00 \times 10^5$ & 0.90 & $1.23 \times 10^4$ & 0.52 & $1.82 \times 10^2$ \\
    5.15h & $1.00 \times 10^5$ & 1.30 & $1.23 \times 10^4$ & 0.75 & $1.82 \times 10^2$ \\
    \hline
    3.1a & $4.10 \times 10^6$ & 13.80 & $5.06 \times 10^5$ & 7.88 & $1.07 \times 10^3$ \\
    3.2a & $7.00 \times 10^5$ & 3.20 & $8.62 \times 10^4$ & 1.83 & $4.59 \times 10^2$ \\
    3.3a & $9.00 \times 10^5$ & 11.40 & $1.11 \times 10^5$ & 6.53 & $5.17 \times 10^2$ \\
    3.1b & $3.60 \times 10^6$ & 15.30 & $4.43 \times 10^5$ & 8.78 & $1.00 \times 10^3$ \\
    3.2b & $5.00 \times 10^5$ & 3.00 & $6.16 \times 10^4$ & 1.72 & $3.91 \times 10^2$ \\
    3.3b & $1.20 \times 10^6$ & 10.90 & $1.48 \times 10^5$ & 6.25 & $5.93 \times 10^2$ \\
    3c & $7.10 \times 10^6$ & 1.40 & $8.74 \times 10^5$ & 0.80 & $1.38 \times 10^3$ \\
    \hline
    1b & $3.90 \times 10^6$ & 6.30 & $4.80 \times 10^5$ & 3.61 & $1.04 \times 10^3$ \\
    2b & $1.10 \times 10^6$ & 6.10 & $1.35 \times 10^5$ & 3.50 & $5.69 \times 10^2$ \\
    3b & $1.01 \times 10^7$ & 24.80 & $1.24 \times 10^6$ & 14.23 & $1.63 \times 10^3$ \\
    4b & $3.10 \times 10^6$ & 4.90 & $3.82 \times 10^5$ & 2.81 & $9.32 \times 10^2$ \\
    5b & $3.30 \times 10^6$ & 8.50 & $4.06 \times 10^5$ & 4.87 & $9.60 \times 10^2$ \\
    \hline
    \end{tabular}
    \tablefoot{
Properties are derived from the cluster data presented in \cite{Vanzella2022a, Vanzella2022b, Vanzella2023}.
}
\end{table}
We now consider the results for the massive young clusters in the Sunburst Arc and other fields covered by \citet{Vanzella2022a, Vanzella2022b, Vanzella2023}. A summary of these clusters is provided in Table~\ref{tab:vanzella_results}. These systems represent the high-mass end of our sample, with several clusters exceeding $10^7 \, {\rm M}_\odot$. While they are spatially extended (half-mass radii often $>10$ pc), the sheer magnitude of their stellar mass still leads to the  production of very massive cores ($M_{\rm core} > 10^6 \, {\rm M}_\odot$), thereby compensating for the more extended radii. The expected distribution of IMBH masses is given in Fig.~\ref{fig:vanzelladistribution},  showing a range of masses from $\sim 1.8 \times 10^2 \, {\rm M}_{\odot}$ up to $\sim 2.1 \times 10^3 \, {\rm M}_{\odot}$.

\subsection{Scaling relations and efficiency}

\begin{figure}[h!]
    \centering
    \includegraphics{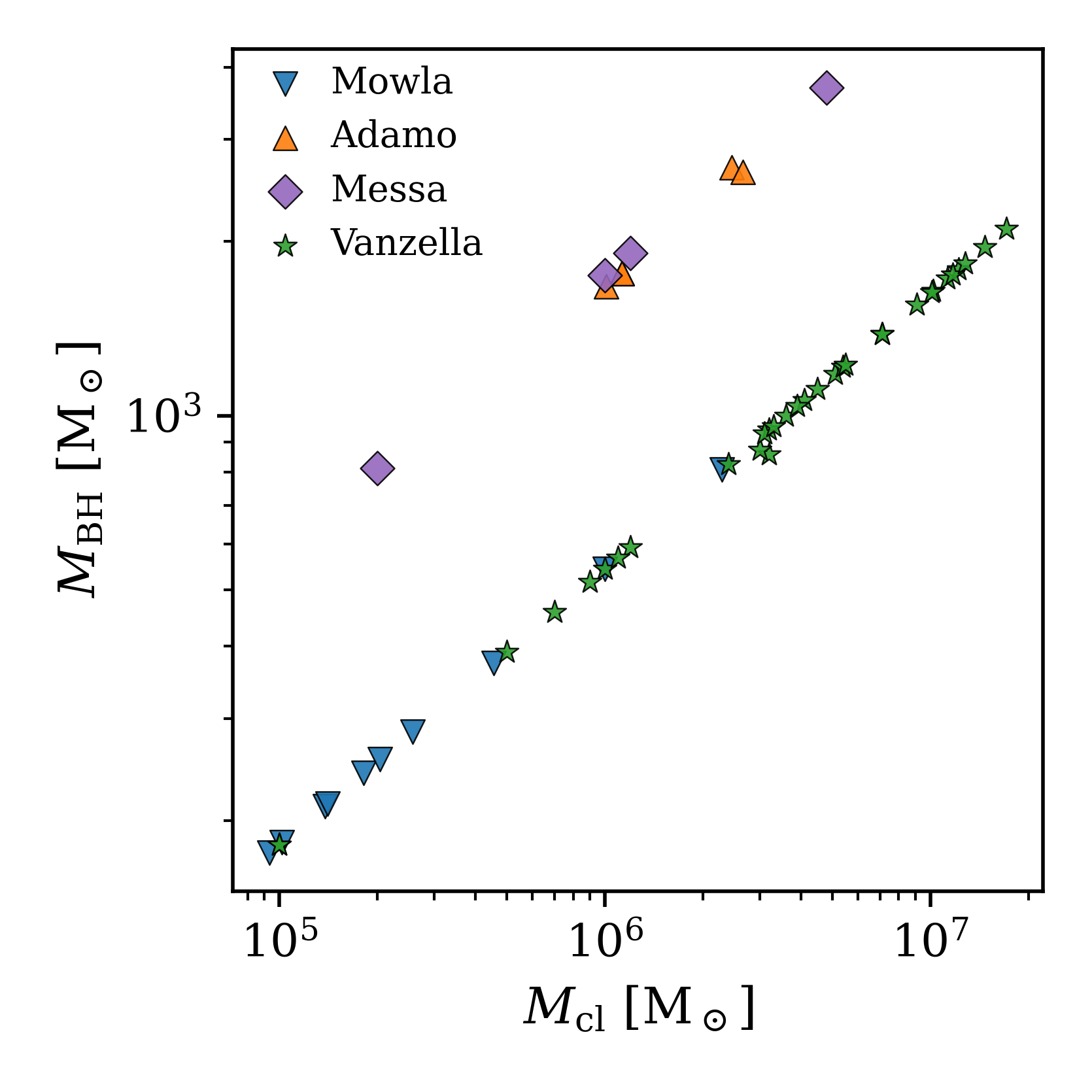}
    \caption{Estimated IMBH mass ($M_{\rm BH}$) as a function of the stellar mass of the host cluster ($M_{\rm cl}$) for the different samples.}
    \label{fig:bhvscore}
\end{figure}

To assess the results more globally, we show the estimated IMBH mass as a function of the core mass for the different samples in Fig.~\ref{fig:bhvscore}. We in general find that most of the dependence of the masses of the IMBHs is due to the mass budget available to go into collisions. A relevant additional factor is also the metallicity. As so far no individual metallicities are provided for each cluster (and their metallicities may also expected to be similar as long as they are in the same environment), we find a good relation between expected black hole mass versus star cluster mass within each given sample, due to the assumption of constant metallicity. On the other hand, the difference in metallicity between the different samples (particularly also the relatively high metallicity case in \citet{Mowla2024}) produces offsets between the different relations. The differences in the radii do lead to fluctuations in these relations but they are not very strong. We do note in this respect that the largest variations in the core radii correspond to a factor of $2-3$, and the mass accretion rate in Eq.~\ref{eq:finalMassAcc} \citep{Portegies2002} depends on the core mass, but not the core radius. We emphasize nonetheless that part of the absence of scatter is also due to the simplifying assumptions in our model, while more scatter could be expected in direct N-body simulations or in case of assuming or considering different cluster properties for example in relation to primordial binaries or the rotation of the clusters. 

For comparison with the work of \citet{Vergara2023, Vergara2024}, we quantify the expected formation efficiency of the IMBHs, which we define as the expected mass of the IMBH divided by the total mass of the cluster. We normalize the cluster mass by the critical mass for runaway collisions, $M_{\rm crit}$, defined as $M_{\rm crit}(R_h) = R_h^\frac{7}{3}\left(\frac{4\pi m_\star}{3\Sigma_0t_{\rm H}G^\frac{1}{2}} \right)^\frac{2}{3}$ \citep{Vergara2023}, where $R_h$ is the radius of the system, $m_\star$ the mass of a single star, $t_{\rm H}$ the age of the system, and $\Sigma_0$  the effective cross section expressed as $\Sigma_0=16\sqrt{\pi}(1+\Theta)\,{\rm R}_\odot^2$, with $\Theta=9.54\left[(m_\star\,{\rm R}_\odot)/(r_\star\, {\rm M}_\odot)\right]\left(100\, {\rm km\,s^{-1}}/\sigma\right)^2$ the Safronov number,where $R_\star$ is the radius of a single star (assumed to be sun-like), and $\sigma=\sqrt{GM_{\rm cl}/R_{\rm h}}$,  which is the velocity dispersion under the assumption of virial equilibrium. The age of the systems is set to $t_{\rm H}=100$\,Myr for all the clusters as a conservative estimate assuming it as the characteristic timescale over which dynamical interactions can operate efficiently as previously discussed in Section \ref{sec:FK}.

\begin{figure}[h!]
    \centering
    \includegraphics{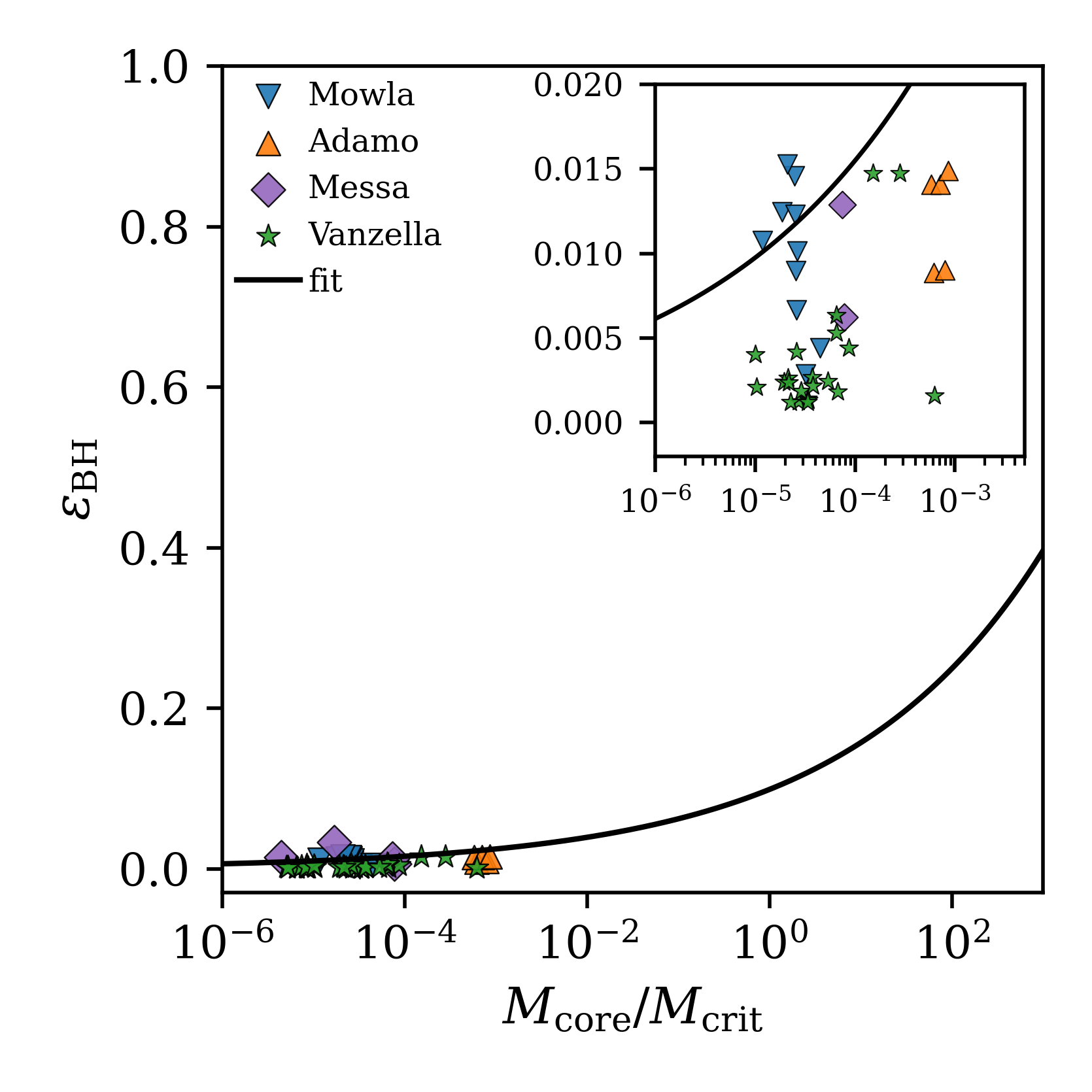}
    \caption{Expected black hole formation efficiency as function of $M/M_{\rm crit}$ from the model presented here and compared to the fit from \citet{Vergara2025b} described in Eq.~\ref{fit}. }
    \label{fig:efficiencyVSMoverMcrit}
\end{figure}

Figure \ref{fig:efficiencyVSMoverMcrit} presents the black hole formation efficiency ($\epsilon_{\rm BH} = [1+M_{\rm BH}/M_{\rm cl}]^{-1}$) as a function of the critical mass ratio. The inset highlights the impact of metallicity. The clusters from \citet{Messa2025} (purple diamonds), which were modeled with higher metallicity ($Z=0.1\, {\rm Z}_\odot$), tend to show lower formation efficiencies compared to similarly massive but metal-poor systems. This suppression is a direct consequence of the metallicity-dependent wind mass loss (Eq. \ref{eq:massLossRate}), which erodes the VMS mass more effectively in chemically enriched environments. For comparison, we also provide the fit from \citet{Vergara2025b} to the numerical simulation data that have explored the dependence of the efficiency parameter $\epsilon_{\rm BH}$ on the ratio $M/M_{\rm crit}$, given as\begin{equation}
\epsilon_{\rm BH}=\left[ 1+\mathrm{exp}\left( -4.63\left[  \log{\left( \frac{M}{M_{\rm crit} }  \right)} -4 \right]\right)\right]^{-0.1}.\label{fit}
\end{equation}
We note that the fit aligns well with the data points including the expected scatter \citep[e.g.,][]{Vergara2024}, suggesting formation efficiencies in the few percent range. This is very reasonable as a result; in fact it seems very likely that clusters with larger formation efficiencies would be much more short-lived and thus very difficult to find via observations.

\section{Summary and discussion}\label{sec:Discussion}

We have estimated the masses of IMBHs formed in the massive young stellar clusters detected by JWST at high redshift using gravitational lensing techniques \citep{Adamo2024, Vanzella2022a, Vanzella2022b,  Vanzella2023, Mowla2024, Messa2025}. As these clusters have large masses up to $10^7$~M$_\odot$, modeling via direct N-body simulations is impossible, and instead we have employed an approximate methodology using Fokker-Planck models to estimate the properties of the central cores as well as the analytical model developed and tested by \citet{Portegies2002} for runaway collisions, extended to consider and include mass loss through winds \citep{Fujii2024, Pacucci2025}, using the wind mass loss rates from \citet{VINK2018}. We have compared the results of this framework to the direct N-body simulations with stellar evolution from \citet{Vergara2025b} for the purpose of verification, finding generally good agreement within a factor of a few, the uncertainty expected within a simplified framework. 

We find from the Fokker-Planck model that the formation of cluster cores with masses in the range from $10^4$~M$_\odot$ up to $10^6$~M$_\odot$ is expected, with core radii in the range from $\sim0.4$~pc up to $\sim15$~pc. Applying the model for the runaway collisions to the different cluster cores, we generally find typical IMBH masses in the range of $10^2-4\times10^3$~M$_\odot$, where the main parameters that regulate the mass of the IMBH are the mass in the core as well as the metallicity, which regulates the mass loss through winds.

For comparison with the results by \citet{Vergara2024, Vergara2025a}, we calculate the efficiency to form a massive object, defined as the mass of the massive object divided by the total mass of the cluster. This quantity is plotted as a function of cluster mass divided by critical mass, where the critical mass is the mass for which the collision time of the system is equal to the evolution time of the system \citep{Vergara2023, Vergara2024}. We compare the results to a fit provided by \citet{Vergara2025b} to direct N-body simulations, finding good agreement, as both methods yield expected efficiencies in the few percent range. This is the range that is also reasonable to expect for observed systems, as the lifetime of stellar clusters may be very short if they are strongly collision dominated. 

We note as a result of these calculations that we can indeed expect the formation of heavy black hole seeds,  with masses $M_{\rm BH} \sim 10^2 -4\times 10^3 \, {\rm M}_\odot$. While the lower end of this mass range is comparable to the $\sim 100 \, {\rm M}_\odot$ remnants expected from Population III stars \citep{Abel2002, Bromm2002}, the upper bound extends significantly higher, allowing the most massive clusters to produce objects firmly in the "heavy seed" regime. Forming such heavy seeds is highly beneficial to form SMBHs at high redshift \citep{Sassano2021}. They can also play a relevant role for gravitational wave emission in the context of IMBH mergers, for example in the context of the Laser Interferometer Space Antenna (LISA)\footnote{Webpage LISA: \url{ https://www.esa.int/Science\textunderscore Exploration/Space\textunderscore Science/LISA}} and the Einstein Telescope\footnote{Webpage Einstein Telescope: \url{ https://www.einstein-telescope.it/en/home-en/}}.

\subsection{Implications for Little Red Dots}
The LRDs discovered by JWST \citep{Matthee2024, Greene2024} exhibit V-shaped spectra often interpreted as broad-line regions powered by SMBHs. Our results suggest that dense massive star clusters could provide possible massive seeds from which the SMBHs in the LRDs may have grown. Particularly,  seeds reaching up to  $3.7\times 10^3 \, {\rm M}_\odot$  (as seen in the most massive cluster) formed at $z \sim 10$ require significantly fewer e-folding times to reach $10^7 - 10^8 \, {\rm M}_\odot$ by $z \sim 6$ compared to light seeds, alleviating the timing constraints on early SMBH growth \citep{Shapiro}. While some scenarios even consider LRDs to be pure stellar systems with very high densities \citep[e.g.,][]{Guia2024}, it can be shown that in such systems, collisions could be expected to be very efficient, leading to the formation of very massive central objects in a similar manner as for the dense massive clusters discussed here, but for a higher mass system \citep{Escala2025, Pacucci2025}. It appears even likely that the LRDs would be in the regime of higher ratios in terms of the mass divided by critical mass, leading potentially to enhanced black hole formation efficiencies in the relation from \citet{Vergara2025b}.

\subsection{Caveats}
The model adopted here is a simplified model that is strongly motivated by our current knowledge on runaway collision processes in stellar systems \citep[e.g.,][]{Portegies2002} and has been tested and compared to the direct N-body simulations by \citet{Vergara2025b}. Nonetheless, the parameters of many of the clusters here are outside of the parameter space that is accessible for direct N-body simulations. We also acknowledge that many of the properties of the clusters are not yet known, including for example the binary fraction, the rotation in the system or the internal structure. The initial mass function is also not known, even if it is possible, if not likely, that we should expect it to be top-heavy \citep{Jevrabkov2018, Kroupa2026}. This possibility, along with a possible primordial mass segregation, would favour the formation of massive IMBHs, though we did not explore it here in detail.

Other important uncertainties to be considered concern the mass loss in the context of stellar collisions. Numerical simulations initially suggested that mass loss to be on the percent level \citep[e.g.,][]{Glebbeek2008, Glebbeek2009}, implying only moderate effects in the context of runaway collisions \citep{Alister2020}. 1D models of stellar pulsation on the other hand suggest that mass loss could also be enhanced, and potentially dominate over the mass gain via collisions \citep{Ramirez2025, Roman2026}. Particularly, the work of \citet{Ramirez2025} shows that the mass loss during collisions depends strongly on the stellar structure and is quite sensitive to the adopted framework for mixing length theory in their calculations, making it crucial to arrive at a better understanding on how to model convection in stars for realistic calculations. The work by \citet{Roman2026} points towards similarly relevant uncertainties and it will be important to understand through future work if the total unbound mass during a collision will be ejected, or if instead their pulsation-based estimate is more accurate. The current range of models within the literature shows a wide range of uncertainties, from minor effects up to potentially strong limitations for VMS formation \citep{Solar2025}.  It will be important for future studies to address this relevant question, including an investigation of pulsations in a 3D framework given the three-dimensional nature of the expected mass loss effects.

\begin{acknowledgements}
We thank the anonymous referee for a careful revision of our paper.
The authors thank for valuable discussions with Marcelo Vergara, Francesco Flammini Dotti, Abbas Askar, Mirek Giersz, Roberto Capuzzo-Dolcetta, Raffaella Schneider, Lorenzo Paparella, Nathan Leigh and Efrain Vira.
DRGS  gratefully acknowledges support from the Alexander von Humboldt - Foundation, Bonn, Germany. ML acknowledges financial support from ANID/DOCTORADO BECAS CHILE 72240058. DRGS thanks for funding via the ANID BASAL project FB21003. 
\end{acknowledgements}

\bibliographystyle{aa} 
\bibliography{astro}

\end{document}